# Naming the Pain in Requirements Engineering: Comparing Practices in Brazil and Germany

Daniel Méndez Fernández[1], Stefan Wagner[2], Marcos Kalinowski[3], André Schekelmann[4], Ahmet Tuzcu[1], Tayana Conte[5], Rodrigo Spinola[6], Rafael Prikladnicki[7]

[1]Technische Universität München, Germany
daniel.mendez@tum.de / ahmet.tuzcu@mytum.de
[2]University of Stuttgart, Germany
stefan.wagner@informatik.uni-stuttgart.de
[3]Universidade Federal Fluminense (UFF)
kalinowski@ic.uff.br
[4]Niederrhein University of Applied Sciences, Germany
andre.schekelmann@hs-niederrhein.de
[5]Universidade Federal do Amazonas (UFAM)
tayana@icomp.ufam.edu.br
[6]Universidade Salvador (UNIFACS)
rodrigo.spinola@pro.unifacs.br
[7]Pontifícia Universidade Católica do Rio Grande do Sul (PUC-RS)
rafael.prikladnicki@pucrs.br

**Requirements Engineering (RE)** constitutes an important success factor for software development projects, since unambiguous and stakeholder-appropriate requirements are critical determinants of quality and productivity [1] [2]. One of the problems we still face, however, is that it is difficult to find proper empirical figures that could demonstrate particular success factors in RE [3]. Empirical studies are inherently difficult for RE with its high variability and long-term feedback cycles. Yet, this makes it difficult to justify the choice and adoption of particular RE practices, as we lack sufficient knowledge about which problems we face in RE, what relevance they have, and what their causes could be. Unfortunately, the lack of evidence in RE leads to the problem that decisions on RE practices are mostly driven and justified by conventional wisdom only.

In response to this problem, we initiated the **Na**ming the **P**ain **i**n **R**equirements **E**ngineering (NaPiRE) initiative[1] hosted under the umbrella of the International Software Engineering Research Network[2]. NaPiRE constitutes a globally distributed family of surveys on the state of RE practices including problems practitioners experience in RE as well as their causes and effects [4]. The resulting knowledge base shall allow us to get a better understanding about the relevance of particular problems and to establish means to mitigate problems in the long run.

Here, we show selected results from the survey edition conducted in 2014/15 by comparing two data sets obtained from two different countries: Brazil and Germany. We intentionally select those countries as their industrial practices show several different characteristics. The data results from responses from 118 Brazilian companies of which we select 74 as they where sufficiently complete, and from 54 German companies of which we select 41 respectively. We use the data to illustrate commonalities and differences in the problems practitioners experience and discuss what we can learn from them. In the following, we directly discuss key results from the survey while a richer description can be taken from our complementary material provided on the project website[1]

I. WHAT IS THE STATUS QUO IN REQUIREMENTS ENGINEERING PRACTICES?

Before discussing the revealed problems in detail, we take a look at selected facets that characterise the practical environments of our respondents (with rounded numbers). Table 1 shows the size distribution of the participating companies for Brazil and Germany. It is possible to observe that our sample includes German companies of larger sizes. 44% of the responding German companies employ more than 2,000 people in contrast to 21% in Brazil. Brazilian responses show with 61% a tendency towards small and medium-sized companies (250 employees and less), which contrasts to Germany with only 29%. In both countries, respondents have at least 3 years of experience in their role (BRA: 75%, GER: 81%). Finally, the organisational role of our responding companies shows a tendency to product development (BRA: 79%, GER: 56%) while the role of a contractor is underrepresented in Brazil (5%) in contrast to our German responses (37%).

---

[1] www.re-survey.org
[2] isern.iese.de

|  | Brazil | Germany |
|---|---|---|
| 1–10 employees | 11 | 3 |
| 11–50 employees | 15 | 6 |
| 51–250 employees | 17 | 3 |
| 251–500 employees | 5 | 2 |
| 501–1,000 employees | 3 | 3 |
| 1,001–2,000 employees | 5 | 6 |
| More than 2,000 employees | 18 | 18 |
| Invalid (missing) answers | 3 | 0 |

Table 1. Size distribution of participating companies.

Some characteristics we see in the data indicate further differences on how projects are conducted and people work together in the countries. The first difference constitutes the chosen process model. In Brazil, we can see a tendency to agile methods such as Scrum (66%) or Extreme Programming (13%), which is slightly higher than in Germany (58% and 2%). This tendency becomes, however, more evident if we consider that German respondents stated to use in 44% of the cases a waterfall model (BRA: 30%), which we can explain by the contracting role where multi-staged bidding procedures often dictate a waterfall-like approach at the beginning of a project. German companies additionally state in 21% of our cases to follow the V-Modell XT in contrast to 5% in Brazil. This is most likely because the V-Modell XT is obligatory for projects in the public sector in Germany. A commonality between both countries is the human-intensive requirements elicitation techniques used. The three most prominent techniques are interviews (BRA: 87%, GER: 88%), facilitated meetings including workshops (BRA: 56%, GER: 86%), and prototyping (BRA: 70%, GER: 52%). One interpretation we have for the differences in facilitated meetings and prototyping is the geographic distribution in Brazil and their slightly higher tendency to use agile methods.

Finally, one striking difference we can observe is that Germany seems to sacrifice standardisation and certification to support the adoption of practices according to individual needs of their project environments. This is reflected in the application of standardised process models (BRA: 54%, GER: 10%) and in the field of RE process improvement, which is dominated in the Brazilian data set by 61% of the cases using external (certifiable) improvement standard while in Germany, 80% of the respondents state to use internal (potentially not certifiable) improvement standards.

II. PROBLEMS! PROBLEMS EVERYWHERE!

The data sets show a variety of problems experienced by our respondents in both countries (see Table 2). If we consider the top 5 problems they experience, they share *Incomplete and/or hidden requirements* as well as *Communication flaws between the project teams and the customer* as common problems. Problems that are experienced differently in both countries are *Underspecified requirements*, *Communication flaws within the project team*, and *Insufficient support by the customer* in case of Brazil and *Time boxing / not enough time, Moving targets, and Stakeholders with difficulties in separating requirements from previously known solution designs* in case of Germany. We interpret this as follows:

- Geographical distribution of companies within Brazil might negatively affect the communication within teams.

- Companies relying more on non-agile, rich process models with a strong contracting component (such as our German respondents) have moving targets as an echoing problem, which could also be a reason for suppressing potential communication problems, which seem more prominent in Brazil.
- Agile methods strongly depend on human-intensive exchange, collaboration, and trust, which, if not apparent, might quickly manifest as problems in a project.

| | Brazil | | | | | | Germany | | | | | | |
|---|---|---|---|---|---|---|---|---|---|---|---|---|---|
| Problem | Prio 1 | Prio 2 | Prio 3 | Prio 4 | Prio 5 | Total | Total | Prio 5 | Prio 4 | Prio 3 | Prio 2 | Prio 1 | Problem |
| Communication flaws between the team and the customer | 9 | 9 | 8 | 3 | 3 | **32** | **20** | 2 | 3 | 2 | 6 | 7 | Moving Targets |
| Incomplete and / or hidden requirements | 12 | 5 | 6 | 5 | 3 | **31** | **18** | 1 | 5 | 2 | 6 | 4 | Incomplete and / or hidden requirements |
| Underspecified requirements | 3 | 14 | 5 | 6 | 3 | **31** | **15** | 2 | 3 | 1 | 3 | 6 | Time boxing / Not enough time |
| Communication flaws within the team | 5 | 5 | 8 | 3 | 5 | **26** | **13** | 0 | 0 | 6 | 4 | 3 | Stakeholders with difficulties in separating requirements from known solutions |
| Insufficient support by customer | 5 | 6 | 6 | 2 | 2 | **21** | **11** | 2 | 1 | 3 | 2 | 3 | Communication flaws between the team and the customer |

**Table 2. Ranking of RE problems in Brazil and Germany with differentiated view on frequency occurrence (Prio 1 to 5).**

III. DOES SIZE MATTER?

We expected to see differences in the experienced problems depending on the company size, because small companies might have a more direct connection to their customers but a less mature process. Therefore, we applied the blocking principle and split the data by company size. We concentrated on two blocks: Large companies with 2000 and more employees and small and medium-sized companies (SME) with 250 employees and less. If we consider large companies only, the diversity in the problems does not change much, but only the details of the problems. However, if we consider SMEs (see Table 3), the mentioned problems become strikingly similar sharing *Incomplete and/or hidden requirements* as the number 1 problem in both countries. Three further problems correspond between the two countries. Only *Insufficient support by customer* appears in Brazil but not in Germany and *Moving targets* in Germany but not in Brazil. Hence, the other problems seem to be much more dependent on the company size than the country context. The occurrence of moving targets in Germany could be explained by the slightly smaller usage of agile methods in Germany. It might also be the case that *Moving targets* and *Insufficient support by customers* are just two results of the communication flaws with the customer.

| | Brazil | | | | | | Germany | | | | | | |
|---|---|---|---|---|---|---|---|---|---|---|---|---|---|
| Problem | Prio 1 | Prio 2 | Prio 3 | Prio 4 | Prio 5 | Total | Total | Prio 5 | Prio 4 | Prio 3 | Prio 2 | Prio 1 | Problem |
| Incomplete and / or hidden requirements | 9 | 3 | 6 | 4 | 1 | **23** | **6** | 1 | 1 | 0 | 3 | 1 | Incomplete and / or hidden requirements |
| Underspecified requirements | 2 | 9 | 2 | 6 | 2 | **21** | **6** | 1 | 1 | 1 | 2 | 1 | Moving targets |
| Communication flaws between the team and the customer | 3 | 5 | 6 | 2 | 1 | **17** | **5** | 0 | 0 | 0 | 2 | 3 | Communication flaws within the team |
| Communication flaws within the | 3 | 4 | 4 | 1 | 4 | **16** | **5** | 0 | 0 | 1 | 1 | 3 | Communication flaws between the |

| | | | | | | | | | | | | |
|---|---|---|---|---|---|---|---|---|---|---|---|---|
| team | | | | | | | | | | | | team and the customer |
| Insufficient support by customer | 5 | 5 | 3 | 1 | 1 | 15 | 5 | 0 | 0 | 4 | 0 | 1 | Underspecified requirements |

Table 3. Top RE problems in small and medium sized companies (up to 250 employees).

IV. IN QUEST FOR A DEEPER EXPLANATION

We were not only interested in identifying problems but also in getting a better explanation for them. For this reason, we analysed the causes and effects of the problems, which might give us indicators on how to mitigate them. Therefore, we asked our respondents for experienced causes and effects as free-text descriptions and coded the answers to make them comparable. For the causes, we used the five suggested categories for causal analysis [5] and for effects, five categories that we found suitable based on the data. In the following, we show the causes and effects of the most striking problem *Incomplete and / or hidden requirements*. Additional relationships between problems and their causes as well as implications is provided in our online material. Figure 1 contains the causes and effects for incomplete requirements in Brazil, the cause effect graph for Germany is shown in Figure 2. The Figures show a probabilistic cause-effect diagram [6], where percentages represent the share of respondents that mentioned each cause or effect.

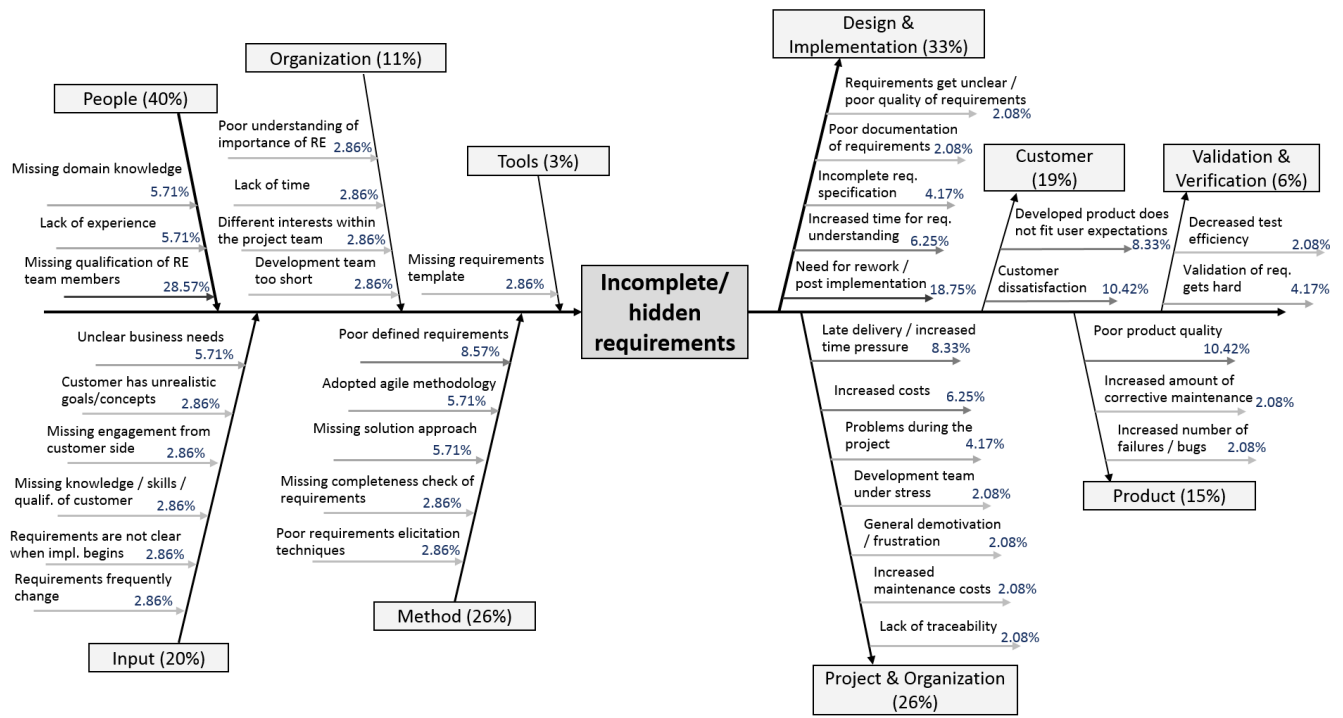

Figure 1. Cause-effect graph for incomplete / hidden requirements in Brazil.

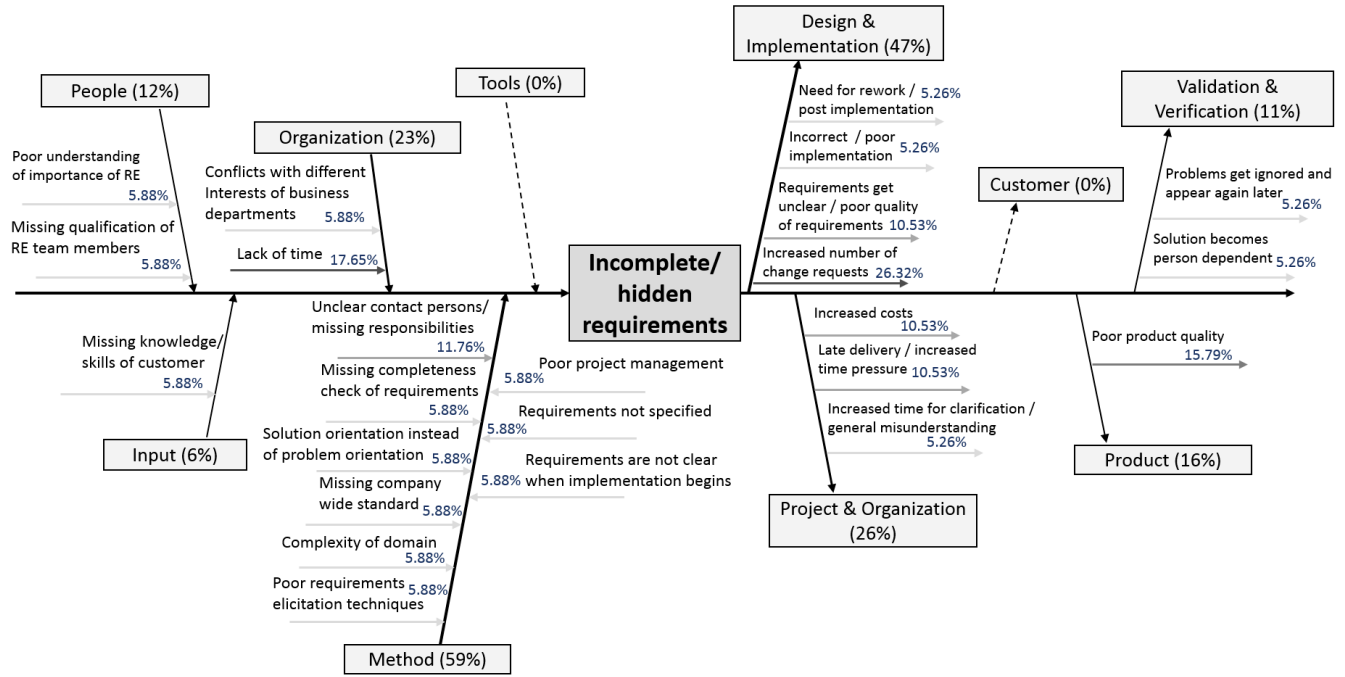

Figure 2. Cause-effect graph for incomplete / hidden requirements in Germany.

We can observe strong differences for this problem in both, causes and effects. The problems experienced in Germany have their causes much more concentrated in the *methods* (59%) and *organisation* (23%) when compared to Brazil, while in Brazil, causes concerning *people* (40%) and *input* (20%) are much more prominent than in Germany. Accordingly, there are also differences in the effects. In Germany, effects on the *Design & Implementation* were with 47% more frequently reported than in Brazil (33%). In Brazil, in turn, effects concerning the *Customer* where present with 19% while in Germany such effects were not reported. We have two interpretations for our observation:

- We recall that in Brazil the companies tend to focus more on product development while in Germany, our respondents also often have the role of a contractor. Hence, the difficulties in gathering input by concrete customers, which might be hard to reach in anonymous markets, might explain the causes and effects in Brazil, while the contractor role in Germany demanding for stronger contracting components might explain the emphasis on the strict methods and the product.

- An alternative interpretation could be a cultural one. The Brazilian respondents might be more aware of problems caused by people issues and with an effect on the customer while the German respondents aim to focus on more objective methods and the product.

## V. WHAT DOES THIS MEAN FOR MY REQUIREMENTS ENGINEERING?

To summarise our observations into one key take-away, we concentrate on one that is a common in our overall data – independent of country, type of projects, or company size: RE is a people's business where human interaction remains key for the elicitation and specification of high quality requirements. This is reflected in the top problems, their causes, and their effects as well.

Our data also indicates that product development motivates companies to certify their processes, which can happen at the cost of adapting methods to individual company cultures. In Germany, we can see that companies individualise their RE (at the cost of certification), which might result in richer process models. Whether this is a cultural aspect or not, the fact is that German companies follow more often rigid processes with stronger contracting components and formal change management. This might also be explained by the company sizes: larger companies feel more the need for specific and formal RE methods while smaller companies need simplicity by agility.

The more the companies rely on agile RE, the less they experience moving targets as a problem. The assumptions on which agile methods rely, i.e. human-intensive exchange, collaboration, and trust, can quickly manifest themselves as severe problems in a project. This is especially true if the communication to the customer is hampered, e.g. by the geographic distribution of a project. We do not yet reliably know whether the manifestation of problems in people issues eventually depends on cultural differences and the awareness for the like, but it seems that more formal process models do not compensate for those problems, but simply shed a different light on them.

**This means,** agility does not necessarily compensate the problems of more formal models, but it can simply make those problems more explicit if a key prerequisite for successful RE is not given: human-intensive exchange, collaboration, and trust.

VI. BIOGRAPHY

Daniel Méndez Fernández is a senior research fellow at the software & systems engineering group at the Technische Universität München. During his PhD and his habilitation, he worked on requirements engineering process improvement and his research interest mainly comprises requirements engineering and empirical software engineering. Further information is available at http://www4.in.tum.de/~mendezfe/

Stefan Wagner is a professor of software engineering at the University of Stuttgart in the Institute of Software Technology. He holds a PhD from the Technische Universität München, an MSc from Heriot-Watt University and a diploma from the University of Applied Sciences Augsburg. His main interests are in quality engineering, requirements engineering and continuous development. Further information is available at http://www.iste.uni-stuttgart.de/se.html

Marcos Kalinowski is a professor of software engineering at the Fluminense Federal University (UFF). He holds PhD and MS degrees from the Federal University of Rio de Janeiro. His research interests mainly comprise empirical software engineering and software quality. Further information is available at http://www.ic.uff.br/~kalinowski

André Schekelmann is a professor of business informatics, esp. software engineering at Niederrhein University of Applied Sciences. He holds PhD and Diploma degrees from University of Paderborn and worked more than 10 years in large-scaled custom software development projects. His research interest is software engineering for business information systems.

Ahmet Tuzcu is a graduate of Technische Universität München with a degree in information systems worked during his master's thesis in the overall evaluation of the NaPiRE study, especially the inference of RE success factors.

Tayana Conte is a professor of software engineering at Universidade Federal do Amazonas (UFAM), where she leads the USES research group (http://uses.icomp.ufam.edu.br/), studying topics such as Usability and User eXperience, learning software organizations and empirical software engineering.

Rodrigo Spínola is a researcher at the Fraunhofer Project Center for Software and Systems Engineering at Federal University of Bahia and a professor of software engineering at the Salvador University where he leads the Technical Debt Research Team (www.tdresearchteam.com). He received his PhD and MS degrees from the Federal University of Rio de Janeiro. His research interests include empirical software engineering with emphasis on maintenance, technical debt, and ubiquitous computing.

Rafael Prikladnicki is a professor of software engineering and director of the Science and Technology Park (TECNOPUC) at Pontifícia Universidade Católica do Rio Grande do Sul (PUCRS), where he leads the MuNDDoS research group (www.inf.pucrs.br/munddos), studying topics such as global software engineering, agile methodologies, crowdsourcing, and high performance software development teams. Contact him at rafaelp@ pucrs.br or via www.inf.pucrs.br/~rafael.

REFERENCES

[1] M. Broy, "Requirements Engineering as a Key to Holistic Software Quality," in *Proc. International Symposium on Computer and Information Sciences (ISCIS)*, 2006, pp. 24-34.

[2] D. Damian and J. Chisan, "An Empirical Study of the Complex Relationships between Requirements Engineering Processes and other Processes that lead to Payoffs in Productivity, Quality, and Risk Management," *IEEE Transactions on Software Engineering*, vol. 32, no. 7, pp. 433–453, 2006.

[3] B.H.C. Cheng, and J.M. Atlee, "Research Directions in Requirements Engineering," in *Proc. Future of Software Engineering (FOSE)*, 2007, pp. 285–303.

[4] D. Méndez Fernández and S. Wagner, "Naming the Pain in Requirments Enginering: A Design for a global Family of Surveys and First Results from Germany," *Information and Software Technology*, vol. 57, pp. 616-643, January, 2015.

[5] M. Kalinowski, D.N. Card and G.H. Travassos, "Evidence-Based Guidelines to Defect Causal Analysis," *IEEE Software*, vol. 29, no. 4, pp. 16-18, 2012.

[6] M. Kalinowski, E. Mendes, and G.H. Travassos, "Automating and Evaluating the Use of Probabilistic Cause-Effect Diagrams to Improve Defect Causal Analysis," in *Proc. International Conference on Product Focused Software Development and Process Improvement (PROFES)*, 2011, pp. 232-246.